\documentclass[fleqn, usenatbib]{mnras}
\usepackage[T1]{fontenc}
\usepackage{ae,aecompl}

\usepackage[]{graphicx}
\usepackage{amsmath}
\usepackage{lmodern}
\usepackage{url}
\usepackage{breqn}
\usepackage{siunitx}
\usepackage{color, soul}
\usepackage{subfigure}

\usepackage{acronym}
\acrodef{WFS}[WFS]{wave-front sensor}
\acrodef{AO}[AO]{Adaptive optics}
\acrodef{ELT}[ELT]{Extremely Large Telescope}
\acrodef{E-ELT}[E-ELT]{European-extremely large telescope}
\acrodef{EST}[EST]{European solar telescope}
\acrodef{FFT}[FFT]{Fast Fourier Transform}
\acrodef{LGS}[LGS]{laser guide star}
\acrodef{MOAO}[MOAO]{multi-object adaptive optics}
\acrodef{DM}[DM]{deformable mirror}
\acrodef{NST}[NST]{new solar telescope}
\acrodef{rms}[rms]{root mean square}
\acrodef{SH-WFS}[SH-WFS]{Shack-Hartmann wavefront sensor}
\acrodef{SST}[SST]{Swedish solar telescope}
\acrodef{SNR}[SNR]{signal to noise ratio}
\acrodef{MCAO}[MCAO]{multi-conjugate adaptive optics}
\acrodef{MOAO}[MOAO]{multi-object adaptive optics}
\acrodef{FOV}[FOV]{field of view}
\acrodef{COM}[COM]{centre of mass}

\makeatletter
\AtBeginDocument{\let\hl\@firstofone}
\makeatother

\graphicspath{{./figs/}}

\title[Reference Images for Correlation Wavefront Sensors]{Generating Artificial Reference Images for Open Loop Correlation Wavefront Sensors}

\author[M. J. Townson, G. D. Love, and C. D. Saunter]{M. J. Townson,$^{1}$\thanks{E-mail: matthew.townson@durham.ac.uk}, G. D. Love,$^{1,2}$ and C. D. Saunter$^{1}$\\
$^{1}$Department of Physics, Durham University, South Road, DH1 3LE, UK\\
$^{2}$Department of Computer Science, Durham University, South Road, DH1 3LE, UK}

\pubyear{2018}

\begin{document}
\label{firstpage}
\pagerange{\pageref{firstpage}--\pageref{lastpage}}
\maketitle

\begin{abstract}
Shack-Hartmann wavefront sensors for both solar and laser guide star adaptive optics (with elongated spots) need to observe extended objects.
Correlation techniques have been successfully employed to measure the wavefront gradient in solar adaptive optics systems and have been proposed for laser guide star  systems. 
In this paper we describe a method for synthesising reference images for correlation Shack-Hartmann wavefront sensors with a larger field of view than individual sub-apertures.
We then show how these supersized reference images can increase the performance of correlation wavefront sensors in regimes where large relative shifts are induced between sub-apertures, such as those observed in open-loop wavefront sensors.
The technique we describe requires no external knowledge outside of the wavefront-sensor images, making it available as an entirely ``software'' upgrade to an existing adaptive optics system.
For solar adaptive optics we show the supersized reference images extend the magnitude of shifts which can be accurately measured from $12\%$ to $50\%$ of the field of view of a sub-aperture and in laser guide star wavefront sensors the magnitude of centroids that can be accurately measured is increased from $12\%$ to $25\%$ of the total field of view of the sub-aperture.
\end{abstract}


\section{Introduction}

\ac{AO} is integral to all next generation solar and \ac{ELT} class ground based telescopes, including the \ac{EST} \citep{Collados2013} and the \ac{ELT} \citep{Davies2010,Cuby2010}.
These telescopes will employ complex \ac{AO} systems which use tomography and multiple \acp{DM} in order to increase either the corrected \ac{FOV} using \ac{MCAO} modalities \citep{Davies2010}, or the number of corrected targets (\ac{MOAO}) \citep{Cuby2010}.

Extended objects are currently the only source available for \acp{WFS} in solar \ac{AO} (typically solar granulation) and are also present on \acp{ELT} due to \ac{LGS} elongation.
The impact of \ac{LGS} elongation on \ac{AO} is the subject of ongoing studies\citep{Thomas2008c,Conan2009,Gilles2006,Schreiber2010a,Anugu2018a}.
\hl{Furthermore, the available} \ac{WFS} \hl{camera options for first light} \ac{ELT} \hl{instruments impose restrictions on the available} \ac{FOV} \hl{resulting in truncation, or limited sampling, of the} \ac{LGS} \hl{plume.}
Similarly, for solar \ac{AO} the effect of how to best use and analyse extended sources in \acp{WFS} is the subject of similar studies \citep{Lofdahl2010}.
One solution to the problem of measuring centroids on extended structure for solar \ac{AO}, proposed by \citet{Rimmele1990}, is the use of correlations in order to accurately measure centroids when observing extended structures.
This technique has been employed successfully in a number of solar \ac{AO} systems, including the \ac{SST} \citep{Scharmer2002}, GREGOR \citep{Soltau2013} and the \ac{NST} \citep{Cao2010} and has also been shown to improve \ac{LGS} \acp{WFS} on-sky \citep{Basden2014}.

Correlation technniques are subject to the same noise sources as other centroiding techniques, including all of the noise sources associated with electrical and optical noise.
However, there is also another source of noise which correlation techniques are subject to, which most other techniques are not.
This extra noise is due to different structure being present in the reference and sub-aperture images which are cross-correlated.
\hl{This noise term is a source of model error, and in the rest of the paper referred to as model error.}
Model error is especially relevant in open-loop \ac{AO} systems, whose \acp{WFS} observe the full strength of atmospheric turbulence, not just residuals after a correction has been applied to the wavefront, as is the case for closed loop \acp{WFS}.

In this paper we propose a computational solution to model error which employs the use of multiple sub-apertures to generate a ``supersized'' reference image.
This supersized reference image can then be used as the reference image in cross-correlations to reduce the model error, as well as other effects which arise from noise in the reference image.
This is shown to improve centoriding accuracy via simulation and allow reliable centroiding for a magnitude of sub-aperture shifts which previously could not be measured.

In order to minimise error sources which are not due to model error the technique described in \citet{Townson2015a} was used in order to choose optimal centroiding parameters for estimating the \ac{COM} of the correlation images.
The simulations were also run in noiseless conditions to only show the effects of model error.

\section{Model Error}
\begin{figure*}
    \centering
    \includegraphics[width=1.\textwidth]{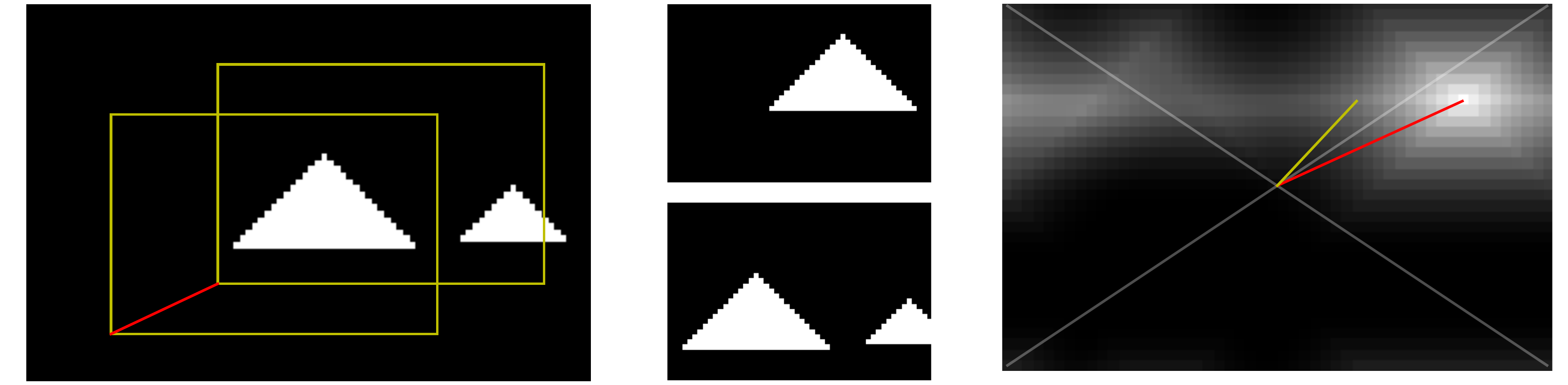}
    \caption{The left part of the figure shows a large, extended, \ac{FOV} which is sampled by a \ac{WFS}.
    The yellow rectangles show the \acp{FOV} of a single sub-aperture and a reference sub-aperture.
    The red line shows the shift between the two images.
    The middle section of the figure shows the sub-aperture image and the reference image individually.
    It can be seen that the sub-aperture image has structure which is not present in the reference image.
    The right part of the figure shows the correlation image from calculating the cross-correlation between the reference image and sub-aperture image.
    The original shift is shown in red and the measured centroid is shown in yellow.
    The centroid is displaced from the true image shift by the extra structure present only in the sub-aperture image.}
    \label{fig:ncin}
\end{figure*}

Model Error is defined here to be noise sources which arise in correlation images from structure which is not present in both the reference and sub-aperture images, \hl{this is also referred to as ``truncation'' for} \acp{LGS} \hl{in night-time} \ac{AO}.
An illustration of this is given in Fig.~\ref{fig:ncin}.
We expect the centroid measurement to correspond to the shift between the sub-aperture image and the reference image.
Here the model error can be seen to add a strong signal to the correlation image, which then skews the centroid measurement away from the value of the shift between the two images.
The left panel shows the ``full'' \ac{FOV}, with the two yellow boxes showing the \acp{FOV} of the reference image and the sub-aperture image.
The red line indicates the shift between the reference and the sub-aperture image.
In the middle panel the reference and sub-aperture images are individually displayed, it can be seen that there is an additional structure in the sub-aperture image which is not present in the reference image.
The right panel shows the resulting correlation image from cross-correlating the reference and sub-aperture images.
The red line shows the shift between the two images and the yellow line shows the measured centroid of the correlation image.
The difference between the centroid and the true image shift is due to model error, and arises from the structure in the correlation image created from the non-common elements between the reference and sub-aperture images.
This shifts the centroid estimate towards it and away from the actual value.

The example shown in Fig.~\ref{fig:ncin} shows a highly simplified case where there are only two structure in the whole field.
For solar granulation there is structure distributed continuously in all directions, this manifests itself as a high background in the correlation image, with many small peaks distributed across it.
In \ac{LGS} images there is continuous structure along the direction of the laser plume.
Although centroiding correlation images which display structure unrelated to the overlap of content common to both images can add noise to the centroid measured there are methods which attempt to minimise the influence of this erroneous structure, such as \citet{Townson2015a,Lofdahl2010}.
These methods make use of windows and threshold values around the peak of the correlation image to remove the influence of non-common structure.

However, while these methods minimise some of the effects of model error they make no attempt to remove it, leaving it present in the correlation images and still making a contribution to the error in any centroid measurement.
Also, for large shifts between sub-aperture and reference images, the overlap region where the peak of the correlation signal is generated is reduced compared to a small shift.
This reduction in the area of overlap for the peak of the correlation image increases the noise on the signal in the correlation image, which in turn adds noise to centroids on the correlation image.
These effects increase when the relative shift between sub-aperture image and the reference image increases, making the problem greater for open-loop \acp{WFS}.

\section{Minimising Model Error}
\label{sec:descriptions}
We suggest that the structure in every sub-aperture image should be wholly contained in the reference image, this should remove the effect of model error.
Sub-aperture images sample different \acp{FOV} due to atmospheric turbulence perturbing the path of incoming wavefronts \citep{Rodd1981,shack1971}.
By combining a set of sub-aperture images, either from a single \ac{WFS} frame or a temporal sample, a larger \ac{FOV} can be reconstructed than the \ac{FOV} contained in any individual sub-aperture image.
This larger image now samples the full \ac{FOV} which all parts of the \ac{WFS} observe.
This generated image can be used as the reference image for correlation \ac{WFS}ing and provide a reference image which includes completely the smaller individual sub-apertures.
The supersized reference image can be used with subsequent frames of a \ac{WFS}, so long as the structure of images in the \ac{FOV} observed remains unchanged.

Each pixel in the generated reference image will be the result of combining multiple sub-aperture images, reducing noise \citep{Basden2014}.
However, for the purposes of this paper we concentrate on the model error aspect of synthesising supersized reference images, especially addressing the ``chicken and egg'' problem of requiring good image shift measurements in order to generate a supersized reference image which in turn is required to measure image shifts.

\begin{figure*}
    \centering
    \includegraphics[width=1.\textwidth]{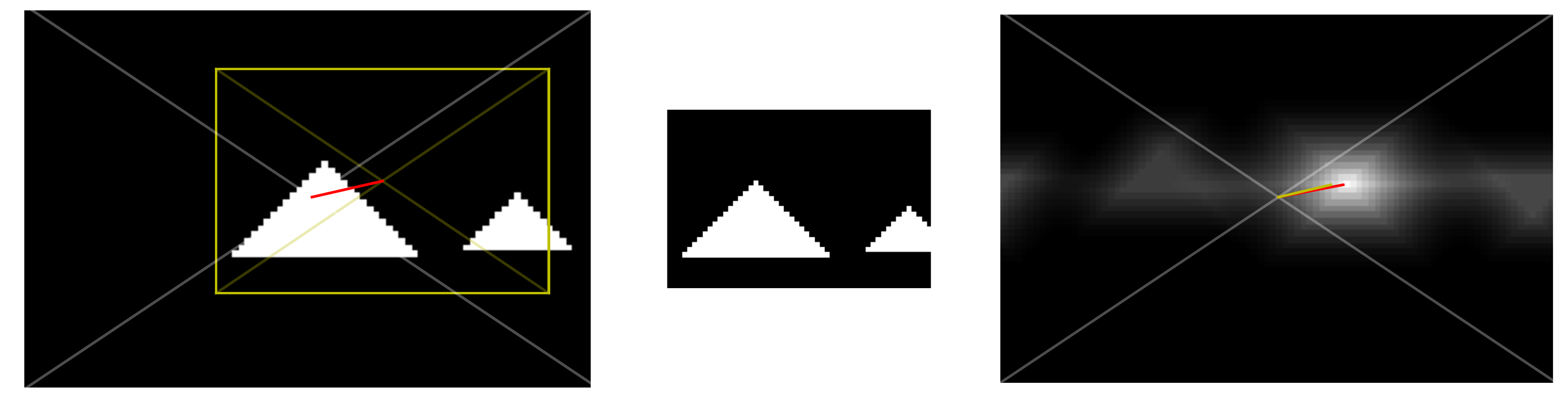}
    \caption{Similarly to Fig.~\ref{fig:ncin} the left part shows a large \ac{FOV}, with the yellow box highlighting the \ac{FOV} of a single sub-aperture.
    The middle region shows just the sub-aperture image, as the reference image is taken to be the full \ac{FOV} shown in the left part of the figure.
    The right part shows the correlation image from a cross-correlation of the sub-aperture image with the full \ac{FOV}.
    The red line indicates the shift of the sub-aperture and the yellow line shows the measured centroid of the correlation image.
    There is still a discrepancy between the shift and the centroid measurement, however, this difference is greatly reduced compared to Fig.~\ref{fig:ncin}.}
    \label{fig:fixed}
\end{figure*}

Figure~\ref{fig:fixed} illustrates the effect of using a supersized reference image in a correlating \ac{WFS} in a similar layout to Fig.~\ref{fig:ncin}.
Here the full field is used as the reference image, so only one sub-aperture \ac{FOV} is highlighted with a yellow window.
When this sub-aperture is correlated with the full field the resulting correlation image shows similar features to Fig.~\ref{fig:ncin}, but the peak of the correlation is much stronger, so the similar structure which adds features to the correlation image have less of an impact on the centroid measurement than it did for Fig.~\ref{fig:ncin}.
This can be seen through the red and yellow lines overlaid on the correlation image, which correspond to the image shift and centroid measurement respectively.
The lines are much closer than they were in Fig.~\ref{fig:ncin}, showing the centroid is more accurate.
\hl{The correlation from the overlap of the larger and smaller triangles is still in the correlation image, biasing the centroid measurement, but it has a reduced effect compared to} Fig.~\ref{fig:ncin}.

In the next section we go on to describe a method to generate supersized reference images using \ac{WFS} data, and show results from simulation on both solar granulation and \ac{LGS} \ac{WFS} images.
However, in order to generate a supersized reference image a set of sub-aperture images must be centroided.
This can be an issue in open-loop correlation \acp{WFS} and generally for correlation \acp{WFS} before the \ac{AO} loop is closed.
Below we describe a method for overcoming this issue.

\subsection{Measuring Large Relative Image Shifts}
\label{sec:large_shifts}
In regimes where the relative shifts between sub-aperture images and reference images is a significant fraction of the \ac{FOV} of the sub-aperture correlation wavefront sensing fails.
\hl{Using a supersized reference image can allow for shifts of this magnitude to be accurately measured, however, there is an issue in generating a supersized reference image in this regime where traditional correlation wavefront sensing fails.}
Also, the centroiding parameters used to measure the location of the peak of a correlation image can add noise \hl{to the measured centroids}.
To minimise noise arising from centroiding parameters we estimate the optimal window and threshold values for a centre of mass on the correlation images using \citet{Townson2015a}.

\begin{figure}
	\centering
    \includegraphics[width=.5\textwidth]{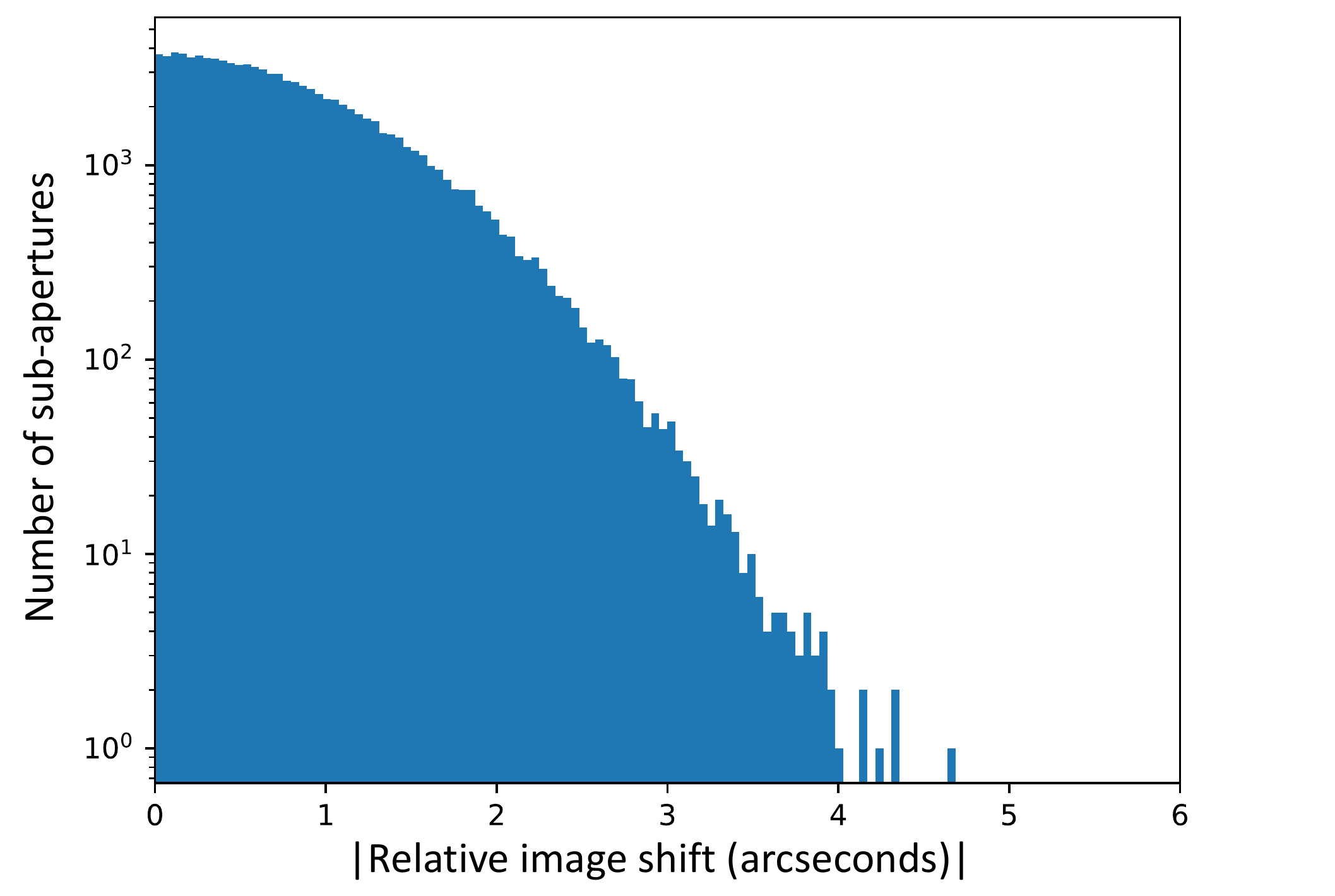} \\
    \includegraphics[width=.5\textwidth]{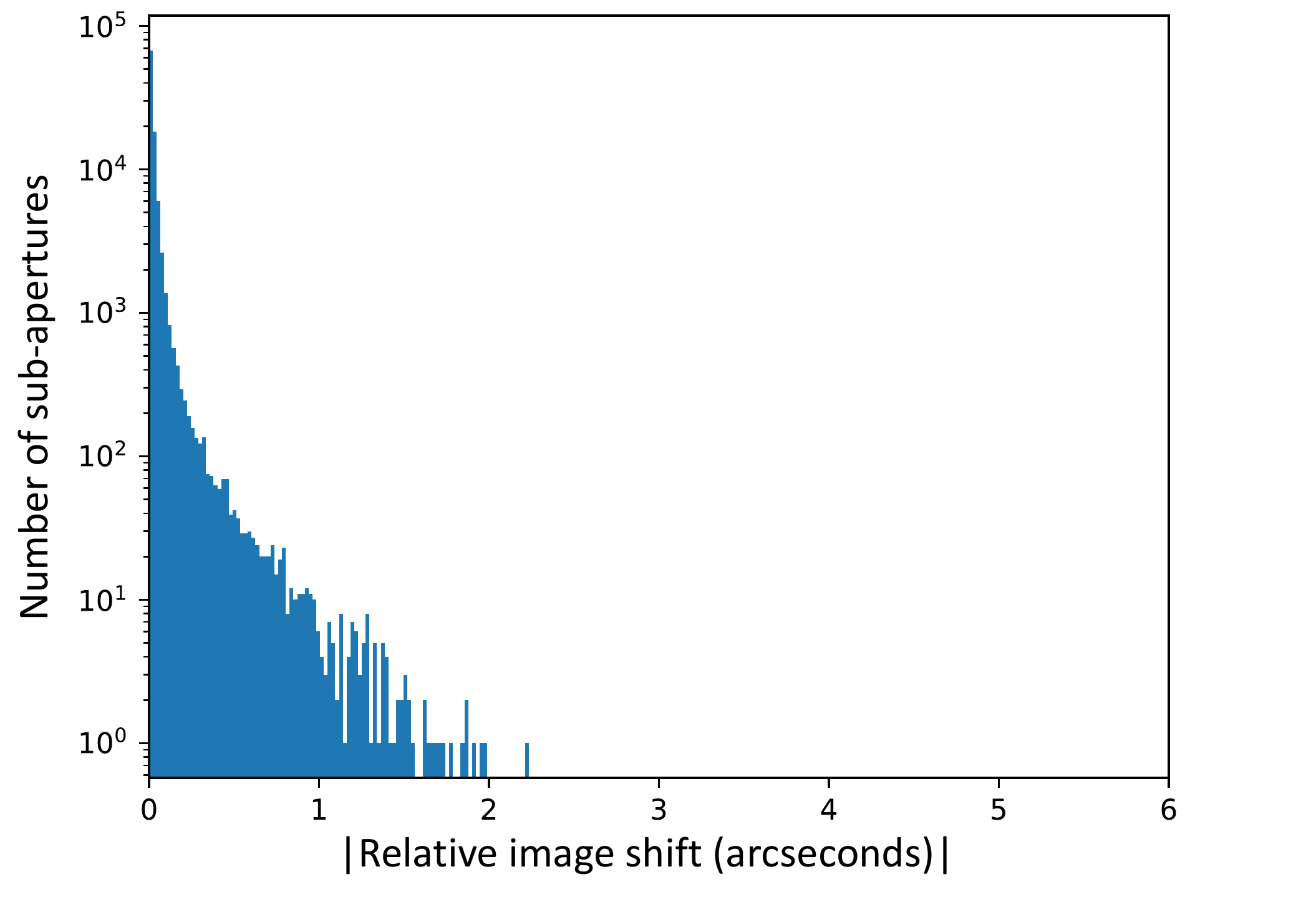}
    \caption{The upper plot shows the distribution in shifts of sub-apertures in a \ac{WFS} for an $r_0$ of $0.10\si{\meter}$.
    For a typical open-loop solar \ac{WFS}, shifts larger than $1\si{\arcsecond}$ are difficult to measure accurately, leaving a significant number of sub-apertures without a reliable centroid.
    However, there is nearly always another sub-aperture within this range in the \ac{WFS}.
    The lower plot shows the absolute smallest relative shift for all sub-apertures in a \ac{WFS}.
    Here there are significantly more pairs of sub-apertures with small relative shifts and a small fraction which have relative shifts above $1\si{\arcsecond}$.
    }
    \label{fig:shift_distribution}
\end{figure}

\begin{figure}
	\centering
	\subfigure[Single reference image]{
		\includegraphics[width=.39\textwidth]{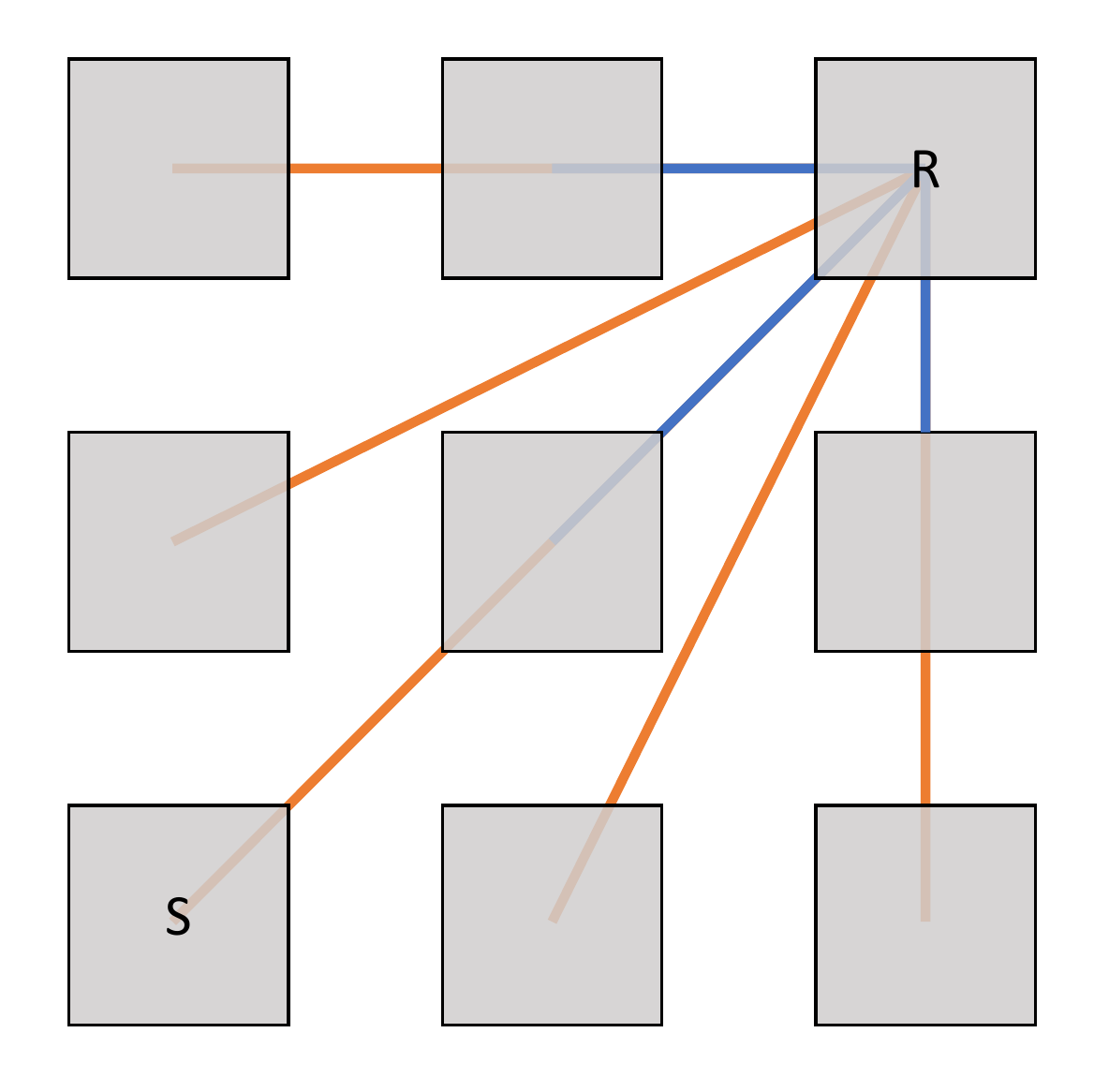}
		\label{fig:a}
	}
	\\
	\subfigure[Tiling reference images]{
		\includegraphics[width=.39\textwidth]{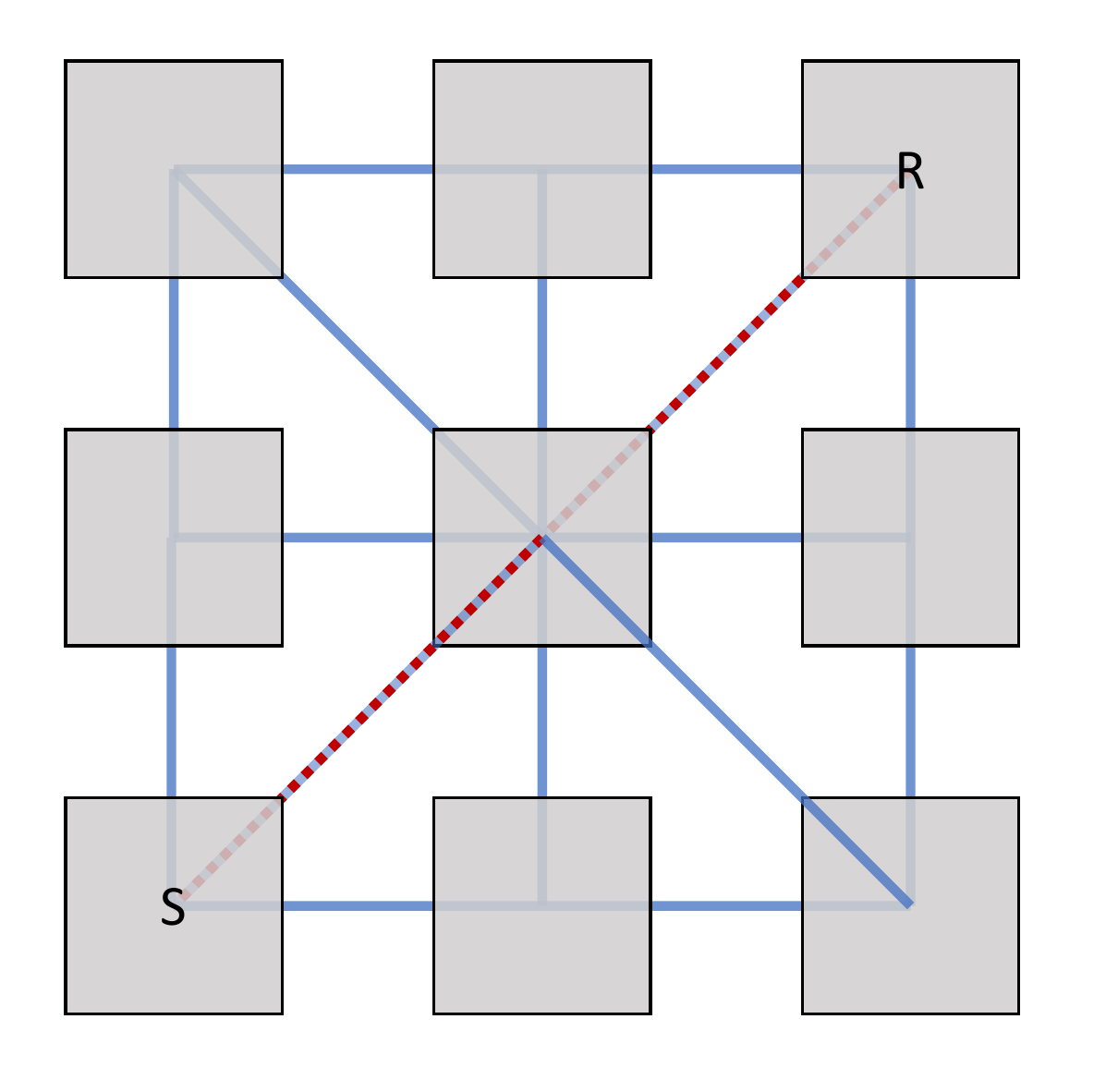}
		\label{fig:b}
	}
	\caption{\hl{A $3\times3$ grid of sub-apertures with cross-correlations shown for a single reference image} \subref{fig:a} \hl{and using all pairs of sub-apertures} \subref{fig:b}.
		\hl{In} \subref{fig:a} \hl{we see three ``good'' cross-correlations, shown in blue, where the sub-apertures are adjacent and have relatively small shifts.}
		\hl{We also see ``bad'' cross-correlations in orange.}
		\hl{These are drawn for sub-apertures which are not adjacent to the reference image, where the model noise is likely to be larger.}
		\subref{fig:b} \hl{shows only the adjacent sub-aperture pairs which are cross-correlated using the `tiling' method.}
		\hl{Whilst there are still many pairs of sub-apertures which are considered to be ``bad'' there are also many which are ``good''.}
		\hl{Every sub-aperture in the } \ac{WFS} \hl{frame can be cross-correlated using only ``good'', blue, cross-correlations.}
        \hl{The absolute shift of each sub-aperture can then can be determined using only information contained in the cross-correlations from adjacent pairs of sub-apertures.}
	}
	\label{fig:lstsq_illustration}
\end{figure}

\hl{Whilst large image shifts exist between some sub-apertures in any particular} \ac{WFS} \hl{frame, adjacent sub-apertures usually have small relative shifts due to the continuous structure of turbulence.}
\hl{Indeed, for any chosen sub-aperture in a} \ac{WFS} \hl{frame there will be a number of other sub-apertures with similar absolute shifts.}
Figure~\ref{fig:shift_distribution} \hl{shows the absolute shift of each sub-aperture for an $r_0$ of $10 \si{\centi\metre}$ in the upper histogram.}
\hl{The lower histogram in} Fig.~\ref{fig:shift_distribution} \hl{shows the smallest relative shift between each sub-aperture and its closest neighbour for a} \ac{WFS} \hl{in the same conditions}.
If we use a pixel scale of $0.25\si{\arcsecond}/\textrm{pixel}$, and a sub-aperture size of 16 pixels, then a shift of $1\si{\arcsecond}$ will be the limit of what can be reliably centroided using a traditional correlation, corresponding to roughly $25\%$ of a sub-aperture width.
\hl{However, by choosing the reference sub-aperture individually for every sub-aperture in the} \ac{WFS} \hl{the measured shift can be minimised, reducing the magnitude of shifts measured from the upper part of} Fig.~\ref{fig:shift_distribution} \hl{to that shown in the lower part.}
The individual relative measurements can then be ``tiled'' across the \ac{WFS} frame and combined to generate the absolute shift of each sub-aperture \hl{as shown in} Fig.~\ref{fig:b}.
\hl{The relative shift between sub-aperture $S$ and $R$ in} Fig.~\ref{fig:a} \hl{is likely to be large as the sub-apertures have a large physical separation, meaning the measured centroid is likely to have a large error.}
In Fig.~\ref{fig:b} \hl{the relative shift between sub-aperture $S$ and sub-aperture $R$ can be found by summing the relative shifts between between adjacent sub-apertures, as depicted by the red dotted line.}
\hl{There are many different combinations of sub-apertures that could be combined to calculate the relative shift between $S$ and $R$, giving many estimates of the shift.}
\hl{These can be combined to reduce the error on the shift measurement.}
This method can be extended to the full \ac{WFS} frame, using a combination of all pairs of sub-apertures to generate correlation images then fitting the individual sub-aperture shifts and using a least squares type.

\hl{Using a least square fit, which includes every possible pair of sub-apertures in a} \ac{WFS} \hl{frame, to estimate the centroids of a set of sub-aperture images mandates every pair of sub-apertures is cross-correlated.}
\hl{While this includes the pairs of sub-apertures with large relative shifts, it also includes all of the larger number of pairs with small relative shifts which can be accurately centroided.}
\hl{This process can be further optimised by adding a weighting function in the least squares fit.}
Centroid measurements which are large can be given a low weighting, and small measures given a greater weighting.
\hl{This suppresses the contribution of pairs of sub-apertures with large measured relative shifts, as these are known to include a larger model error.}
This leaves us with the least squares problem of the form described in Eqn.~\ref{eqn:lst_squares};
\begin{equation}
  \left( \begin{array}{ccc}
    w^{0,1} & -w^{0,1} \\
    w^{0,2} & & -w^{0,2} \\
    & w^{1,2} & -w^{1,2} \end{array}
  \right)
  \left( \begin{array}{c}
    \mathbf{R}^0 \\
    \mathbf{R}^1 \\
    \mathbf{R}^2 \end{array}
  \right)  
   = \left( \begin{array}{ccc}
      w^{0,1} \\
      w^{0,2} \\
      w^{1,2} \end{array}
  \right)
  \left( \begin{array}{ccc}
      \mathbf{R}^{0,1} \\
      \mathbf{R}^{0,2} \\
      \mathbf{R}^{1,2} \end{array}
  \right)^T,
\label{eqn:lst_squares}
\end{equation}
where $w^{i,j}$ is the weight applied to the centroid of sub-aperture $i$ cross-correlated with sub-aperture $j$, $\mathbf{R}^i$ is the absolute shift of sub-aperture $i$ and $\mathbf{R}^{i,j}$ is the relative shift between sub-apertures $i$ and $j$.
The values of $\mathbf{R}^{i,j}$ are measured by centroiding the correlation image between the sub-apertures and the weights applied are the inverse of the magnitude of these centroids, such that;
\begin{equation}
	w^{i,j} = \frac{1}{|\mathbf{R}^{i,j}|}.
	\label{eqn:weight}
\end{equation}
There are other metrics which could be used as a weighting variable for the least squares fit, such as the contrast or sharpness of the individual sub-aperture images or the residuals from a unitary weighted fit.
However, using the magnitude of the measured shift is a natural solution to the problem we are trying to solve in that it suppresses the measurements which experience the largest model error.
On real \ac{AO} systems there may be other noise effects which mean a different choice of weighting function would be more appropriate.

\subsection{Reference Image Generation}
Using the technique described in the Sec.~\ref{sec:large_shifts} a series of sub-aperture images, with large relative shifts, from a \ac{WFS} can have their individual image shifts measured.
The simplest way to synthesise a reference image from \ac{WFS} images and image shift measurements is to align the individual images using the shift measurements and sum them.
This has been implemented before by \citet{Basden2014}, where the technique is used to increase the \ac{SNR} of the reference image for \ac{LGS}.
However, by keeping all parts of the aligned image rather than cropping to the \ac{FOV} of a sub-aperture, we can create a supersized reference image. 
This has the effect of reducing noise in the reference image, in a similar way to \citet{Basden2014}, but also offers a solution to model error.

By using all sub-apertures in a \ac{WFS} frame a supersized reference image can be created which contains no regions which are not sampled by at least one sub-aperture image.
The fill factor can be increased further if multiple frames are used together to create a supersized reference iamge.
However, the edges of the supersized reference image can still be under-sampled if there are too few sub-aperture images with large absolute shifts.

\begin{figure}
    \centering
    \includegraphics[width=\columnwidth]{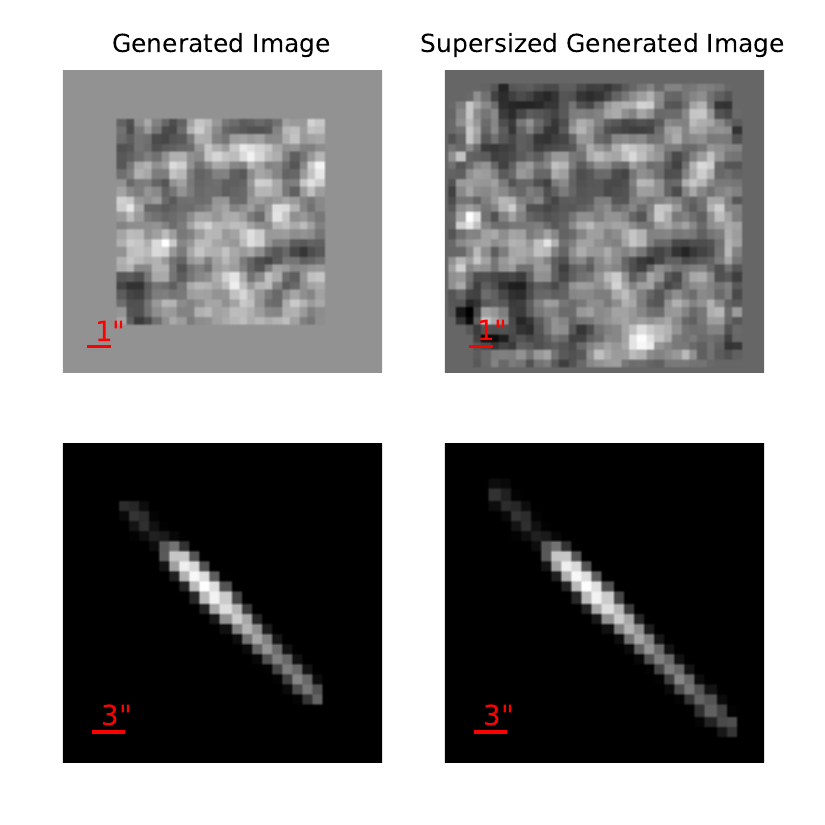}
    \caption{\label{fig:supersize_im}
    		Images of generated reference images shown beside supersized reference images.
            The upper images show the case for images of solar granulation and the lower images show the case for images of elongated \acp{LGS}, shown on a logarithmic scale to highlight the edges of the laser plume.
            The left images show the full \ac{FOV} observed by a single sub-aperture and the right images show the full synthesised supersized reference images.
        The scale of the sub-aperture and supersized reference images is the same for each object, with the padding showing the areas where extra information present in the supersized reference images.
            }
\end{figure}

Here, to create supersized reference images initially the sub-aperture images were up-sampled to $10\times$ their original size, to enable alignment of the images to a sub-pixel scale.
The images were then stacked, rounding the shifts to the nearest tenth of a pixel, so the shifts were all an integer number of pixels on the up-sampled sub-aperture images.
All parts of the images were kept after the alignment, such that the resulting stacked image had a larger \ac{FOV} than the original \ac{FOV} in the sub-aperture images.
Although \ac{FFT} methods exist for co-aligning images, these were not employed as they typically do not allow for a larger sized output image than input.
This super-sized image was then binned back down to the original scale of the \ac{WFS}, generating an image similar to the input sub-aperture images, but with higher \ac{SNR} and a larger \ac{FOV}.
Example synthesised supersized reference images from simulated solar and \ac{LGS} \ac{WFS} images are shown in Fig.~\ref{fig:supersize_im}.
The \ac{LGS} image contains elongation which is representative of those expected to be observed at the edge of the \ac{ELT} pupil.

The initial supersized reference images which are created from using the centroids from the least squares fit described in Sec.~\ref{sec:large_shifts} are not always sufficient be used as reference images for cross-correlation.
This is due to the accuracy of the centroid measurements which arise from the least squares technique.
However, the synthesised supersized reference image does contain a larger \ac{FOV} and show the structure found in the sub-aperture images, so can be used in order to estimate the shifts of the same set of sub-apertures again, to a higher accuracy.
This process can then be repeated, in a bootstrap type method, multiple times in order to ``refine'' the synthesised supersized reference image.
After a certain number of iterations the centroid estimates will no longer change from each iteration, so the generated supersized reference image stabilises.
This occurs when the centroids from the sub-apertures is optimal, so the synthesised reference image is also optimal for the input set of sub-aperture images.
This supersized reference image can then be used as the reference image to measure the centroids in subsequent \ac{WFS} frames.

Due to the distribution of shifts in Fig.~\ref{fig:shift_distribution} the edges of the supersized image is contained in relatively few of the sub-aperture images.
The resulting \ac{SNR} in a supersized reference image therefore varies across the supersized image.
In the centre of the image the \ac{SNR} is highest and it decreases further from the centre.
The resulting shift estimates from using a supersized reference image will therefore be more reliable for small shifts and less reliable for large shifts, where the edges of the supersized reference images contribute signal to the cross-correlation.
This effect can be reduced by using more sub-aperture images, from different frames, to generate a supersized reference image.
Though the estimates from a supersized reference image with poor \ac{SNR} at the edges of the \ac{FOV} will be more accurate than if there was no structure in the reference image {\it i.e.} the reference image had a smaller \ac{FOV}.
This effect of differential noise properties across the supersized reference image is not investigated here, as noiseless images are used in the simulations to only show the effect of model error.

\section{Comparison of Reference Images}
\begin{figure}
    \centering
    \includegraphics[width=\columnwidth]{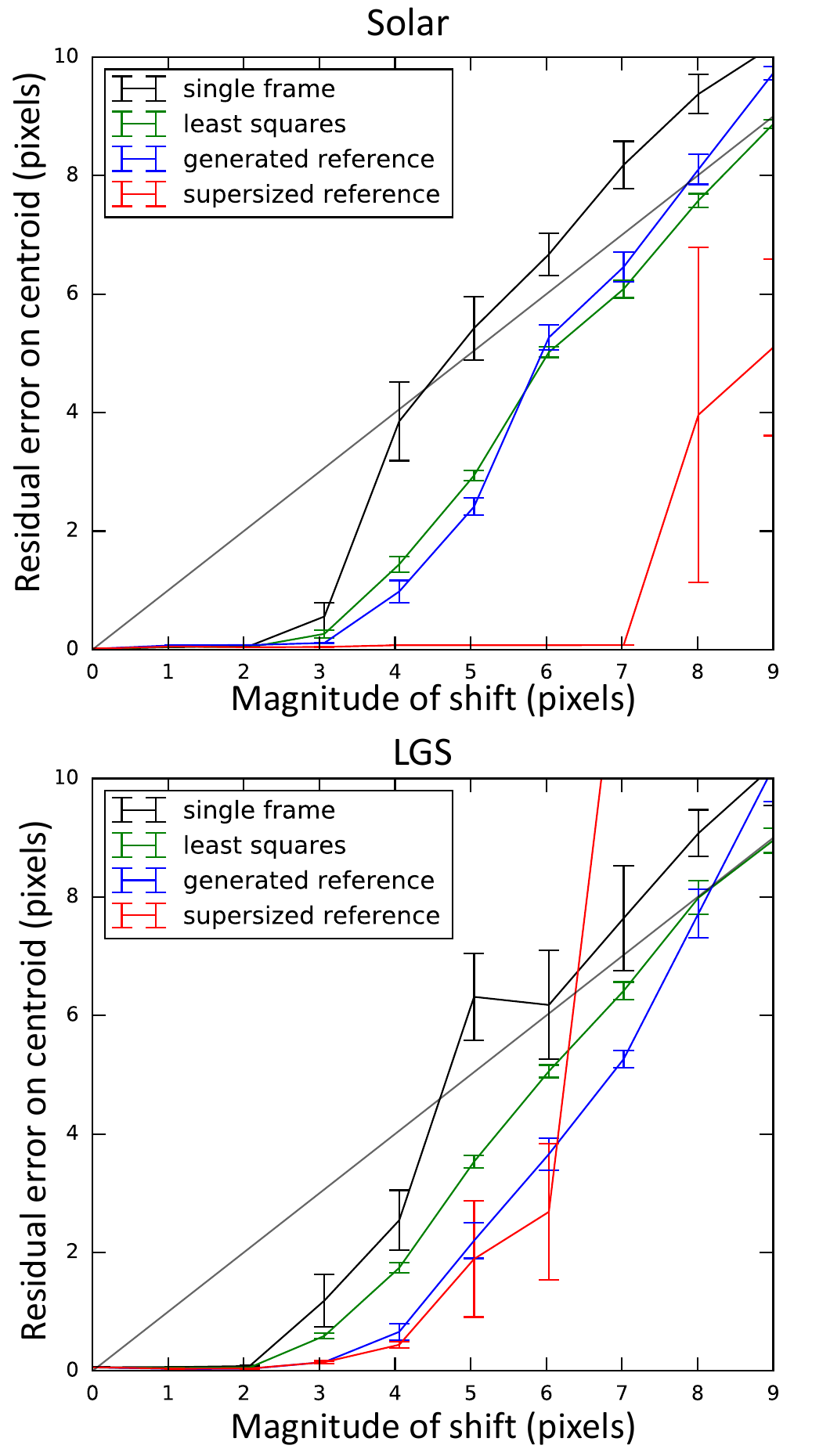}
    \caption{\label{fig:superref}
    The upper figure shows the results of the various correlation centroiding techniques on images of solar granulation and the lower plot shows the same techniques applied to highly elongated \ac{LGS} images.
    The black line shows the residual error from using a single sub-aperture image as the reference image in a cross-correlation.
    The green line shows the performance of the least squares technique described in sec.~\ref{sec:large_shifts}, the blue line shows the performance of a synthesised reference image windowed to the same \ac{FOV} as a single sub-aperture and the red line shows the performance of the full supersized reference image.
    The grey line shows $y=x$ for reference of centroiding accuracy.
    All centroiding methods perform similarly, better than 0.1 pixels, for shifts of less than 3 pixels.}
\end{figure}

The technique described in Sec.~\ref{sec:descriptions} was compared with the widely used method of using a single sub-aperture image as the reference image in correlation \ac{WFS}ing.
\ac{WFS} images of solar granulation and \acp{LGS} were simulated with relative shifts representative of Von-Karman atmospheric turbulence, and centroided by cross-correlating them with a single, central un-vignetted sub-aperture as the reference image in order to obtain a baseline performance for the standard method of performing correlation \ac{WFS}ing.
The least squares method of measuring centroids, described in Sec.~\ref{sec:large_shifts} was performed and its \ac{rms} error measured, along with the \ac{rms} error from using synthesised reference images generated from the \ac{WFS} image.
The synthesised reference images were used in two situations, the first where the supersized reference image was windowed to the original \ac{FOV} of a single sub-aperture, comparable to the method used in \citet{Basden2014}, and the second where the full \ac{FOV} of the supersized reference image was used.
The results for both sub-aperture images of solar granulation and highly elongated \ac{LGS} images are shown in Fig.~\ref{fig:superref}.

Using a single sub-aperture as the reference image works well in both the solar and \ac{LGS} case for small shifts, of up to ~2 pixels ($12.5\%$ shift of \ac{FOV}).
This is expected as the simulations were run in noiseless conditions, so where there is a large overlap of sub-aperture images with the reference image the centroid estimate taken from using a cross-correlation has a high \ac{SNR}.

The initial method for estimating the centroids for creating a supersized reference image, using the least squares method described in Sec.~\ref{sec:large_shifts}, performs better than simply using a single sub-aperture as a reference image.
It can be seen in both the solar and \ac{LGS} case to consistently outperform using a single sub-aperture image as a reference, and also crucially falls below the $y=x$ grey line for very large shifts of up to $50\%$ of the total \ac{FOV} of the sub-apertures.
This allows the bootstrapping technique of iteratively using generated reference images on the same set of sub-aperture images to refine the shift estimates to work.
However, alone the least squares technique offers very little advantage when operating an \ac{AO} system.
This is due to the fact the error on the centroid measurements is larger than $\sim 0.1 \textrm{pixel}$ at the same magnitude of image shifts as a single sub-aperture used as the reference image.

After generating a supersized reference image, if a windowed version is used, such that it has the same \ac{FOV} as the sub-aperture images, the accuracy of centroiding is not improved for the case of daytime observation compared to the least squares method, but there is a less significant improvement in centroiding accuracy for the \ac{LGS} case.
The difference in performance is due to the general structure typically observed for each of the different cases and is explored in more detail in Sec.~\ref{sec:comparison}.
Also, for noiseless simulations it is interesting that the generated reference offers better performance than using a single sub-aperture as the reference.
This is due to the statistics of the image shifts, as the generated reference image will always be centered on an image shift of 0 pixels, whereas the absolute shift of a single sub-aperture can be anywhere in the distribution, skewing the relative shifts with respect to a single sub-aperture used as the reference image.

The full supersized reference image performs the best of all the techniques shown here.
For small shifts the supersized reference image performs at a similar level to the other methods, and then offers a significant increase in performance for increasingly large shifts until it begins to fail at shifts with a magnitude of 7 pixels for images of solar granulation.
This accuracy in centroiding measurements remains at the sub-pixel level for all magnitudes of shifts for the sub-aperture images until the shifts reach $50\%$ of the sub-aperture \ac{FOV}, a much greater shift than any of the other methods are able to perform up to.
For the \ac{LGS} case the supersized reference offers a much more modest increase in centroiding accuracy and is already struggling to achieve a performance required for \ac{WFS}ing in \ac{AO} when the magnitude of shift reaches $25\%$ of the sub-aperture \ac{FOV}.
This discrepancy in performance between the two types of images is due to the difference in general structure contained in the images and is discussed more below.

\subsection{Comparison of Solar and Laser Guide Star Reference Images}
\label{sec:comparison}
The main differences in performance between the solar granulation images and the images of a \ac{LGS} shown in Fig.~\ref{fig:superref} are in the performance of the synthesised reference images.
The images of solar granulation contain a continuous structure across the \ac{FOV} in all directions, whereas the images of \acp{LGS} only contain structure along one dimension (the direction of elongation).
This restricts the advantage of using a supersized reference image to one dimension for the \ac{LGS}, making the gains of the supersized reference image better for the solar case than for the \ac{LGS} case.

This differing structure also affects the relative performance from using a windowed synthesised reference image from combining multiple sub-aperture images.
The windowed generated reference image is more effective for solar granulation as the structure in the sub-aperture images extends over all directions.

Another difference with \ac{WFS} frames observing \acp{LGS} arises from the fact that the orientation and level of elongation seen in a sub-aperture is dependant on the exact geometry and location of the sub-aperture, which differs between sub-apertures.
This can be mitigated to a certain extent by taking a time series of images from a single sub-aperture and synthesising a reference image for every sub-aperture from a temporal set of frames from the same sub-aperture.
The generated reference images would then vary across pupil for different sub-apertures.
However, this assumes that the observed Sodium plume is stable over the time period the data set for creating the supersized reference image as well as the length of time the synthesised reference image is the applied for.
This assumption seems reasonable, with data from \citet{Pfrommer2010a, Michaille2001} suggesting the Sodium layer is stable over periods of minutes, apart from occasional spikes in the profile from micro meteorites, at a rate of $\sim20\si{\per\hour}$.

\section{Discussion}
Overall, using a supersized reference image in a correlation \ac{WFS} offers an advantage over other types of reference images, which are restricted to the \ac{FOV} of a single sub-aperture image.
Supersized reference images offer improved performance where shifts between sub-aperture images are large (over $12\%$ of the sub-aperture \ac{FOV}).
This is due to the fact that using a supersized reference image combats model error in the correlation images.
In the solar case, where structure is continuous in all directions around a target \ac{FOV}, the technique offers a level of performance similar to standard correlation centroiding techniques which observe small shifts, but extend the magnitude of shift which can be reliably centroided from $\sim 12 \%$ to $\sim 50\%$ of the \ac{FOV} of a sub-aperture.
Above this magnitude of shift a failure region is reached.
This region is determined by the ability of reliable centroid estimates to be made initially, which can then be iterated upon to generate the final supersized reference image.
The technique could work with a larger magnitude of shifts if the initial estimates of centroids were improved.
This technique has direct implications for existing open-loop \ac{WFS} instruments, such as S-DIMM+\citep{Scharmer2010} and Solar SLODAR\citep{Townson2016a}.

The technique performs less well for sub-apertures observing highly elongated \ac{LGS}.
\hl{Previous studies have shown cross-correlation and matched-filter to offer better centroid accuracy than simply centroiding the} \ac{LGS} \hl{plume} \citep{Basden2014,Basden2017}.
This is due to the signal being restricted to one dimension, along the direction of elongation.
However, an improvement in the accuracy of centroid estimates is still observed, until like for the solar case a ``catastrophic'' failure point is reached.
The technique does offer a valid improvement to the accuracy of centroid estimates, achieving a sub-pixel accuracy for \acp{FOV} of up to $25\%$ of the total \ac{FOV}, rather than the $12\%$, which conventional reference images offer.
\hl{This is due to the larger} \ac{FOV} \hl{in the reference image reducing the impact of truncation on the centroiding measurements.} 
Generating supersized reference images also does not require any external knowledge of the Sodium layer, so no external observations of the Sodium plume are required.

There is a significant overhead associated with initially generating a supersized reference image as described here.
The least squares method requires significant amounts of computation to generate centroids, then subsequent iterations to improve the centroid estimates for all pairs of sub-apertures with successive ``generations'' of synthesised supersized reference images are required.

However, this process is only required to create an initial supersized reference image, so only performed once.
The initial step to generate a first supersized reference image scales as the number of sub-apertures in the \ac{WFS}, \hl{with the total number of correlations required scaling as $O(n^2)$ for $n$ sub-apertures, where using a single reference image scales as $O(n)$.}
Even for \ac{ELT} scale \ac{LGS} \ac{WFS} ($80 \times 80$ sub-apertures) a supersized reference could be generated in $\sim 1 \si{\second}$ on the \ac{WFS} processing hardware.

Updating a supersized reference image, because either the structure in the \ac{FOV} has evolved or the conditions have changed significantly, can be performed using the images of a previous \ac{WFS} frame, or temporal set of images, and the accurate centroids from the working \ac{AO} system.
Utilising the output centroids from a working \ac{AO} system eliminates the overhead associated with measuring centroids for generating the supersized reference image, vastly reducing the computational cost in a working system.
Updating the reference image would add negligible computational overhead to the \ac{AO} system, \hl{as the centroiding is already performed as part of the} \ac{AO} loop.
Running an \ac{AO} \hl{loop with a supersized reference image increases the computational cost of centroiding compared to using a single sub-aperture as a reference image.}
\hl{For a typical solar} \ac{WFS} \hl{with $16 \times 16$,} $0.25 \si{\arcsecond}/\textrm{pixel}$ \hl{pixels in a sub-aperture in conditions with $\textrm{r}_0=10\si{\centi\meter}$ the supersized reference image would typically be $20\times20$pixels, an increase of $0.25\times$.}
T\hl{his increase in size increases the centroiding complexity by $1.6\times$.}
\hl{However, in the full} \ac{AO} \hl{loop this increase in computation takes centroiding from $\sim10\%$ of the total computation time to $\sim14\%$ of the total computation time, an increase of $<5\%$.}

\hl{Increasing the magnitude of tilt a} \ac{WFS} \hl{can measure allows for a number of avenues to be explored.}
The technique could be used for closed-loop \ac{WFS} design.
By improving the magnitude of shift that can be measured in a given \ac{WFS}, finer sampling of the \ac{WFS} target could be used on a reduced \ac{FOV}.
This would potentially allow for more accurate centroids to be attained in closed-loop \ac{WFS}, or to circumvent issues which arise from the limited number of pixels available in fast, sensitive cameras which are required for \ac{WFS}ing, \hl{such as the truncation of} \ac{LGS} \hl{images}.
This is highly relevant for the case of \ac{LGS} \ac{WFS} for \acp{ELT}, where there is a trade-off between the number of sub-apertures in the \ac{WFS} and the sampling resolution of the \ac{LGS}.
\hl{Using a smaller} \ac{FOV} \hl{in Solar} \acp{WFS} \hl{would also increase the sensitivity of the} \ac{WFS} \hl{to high altitude turbulence.}

\section*{Acknowledgements}
MJT gratefully acknowledge support from the Science and Technology Facilities Council (STFC) in the form of a PhD studentship (ST/K501979/1) and grant (ST/P000541/1).
The authors would like to thank the Institute of Solar Physics, Sweden, Mats Carlsson, Viggo Hansteen, Luc Rouppe van der Voort, Astrid Fossum and Elin Marthinussen for taking the raw solar image used in this paper, and Mats L\"{o}fdahl for performing the image reconstruction to produce the final solar image.
Data used are available from the author on request.

\bibliographystyle{mnras}
\bibliography{Mendeley.bib}

\bsp	
\label{lastpage}
\end{document}